# CleanCTG: A Deep Learning Model for Multi-Artefact Detection and Reconstruction in Cardiotocography

Sheng Wong[a], Beth Albert[a], Gabriel Davis Jones[a]

a. Oxford Digital Health Labs, Nuffield Department of Women's & Reproductive Health, University of Oxford, Women's Centre, John Radcliffe Hospital, Oxford, UK

Corresponding author: Gabriel Davis Jones

## Abstract

Cardiotocography (CTG) is essential for fetal monitoring but is frequently compromised by diverse artefacts, including beat halving and doubling, maternal heart rate overlap, missing segments and isolated spikes, which obscure true fetal heart rate (FHR) patterns and can lead to misdiagnosis or delayed intervention. Current deep-learning approaches typically bypass comprehensive noise handling, applying minimal preprocessing or focusing solely on downstream classification, while traditional methods rely on simple interpolation or rule-based filtering that addresses only missing samples and fail to correct complex artefact types. We present CleanCTG, an end-to-end dual-stage model that first identifies multiple artefact types via multi-scale convolution and context-aware cross-attention, then reconstructs corrupted segments through artefact-specific correction branches. Training utilised over 800,000 minutes of physiologically realistic, synthetically corrupted CTGs derived from expert-verified "clean" recordings. On synthetic data, CleanCTG achieved perfect artefact detection (AU-ROC = 1.00) and reduced mean squared error (MSE) on corrupted segments to $2.74 \times 10^{-4}$ (clean-segment MSE = $2.40 \times 10^{-6}$), outperforming the next best method by more than 60%. External validation on 10,190 minutes of clinician-annotated segments yielded AU-ROC = 0.95 (sensitivity = 83.44%, specificity 94.22%), surpassing six comparator classifiers. Finally, when integrated with the Dawes-Redman™ system on 933 clinical CTG recordings, denoised traces increased specificity (from 80.70% to 82.70%) and shortened median time to decision by 33%. These findings suggest that explicit artefact removal and signal reconstruction can both maintain diagnostic accuracy and enable shorter monitoring sessions, offering a practical route to more reliable CTG interpretation.

# Highlights

- We introduce CleanCTG for noise detection and denoising in CTG, trained on over 800,000 minutes of synthetic, physiologically realistic CTG.
- CleanCTG outperformed other models in all evaluation stages.
- CleanCTG achieved an AUC of 0.95 in noise detection and MSE of $2.74 \times 10^{-4}$ in denoising corrupted CTGs.
- Integration with the Dawes-Redman™ system demonstrated significant clinical utility, achieving a median 33% improvement in time to decision.

# Introduction

Cardiotocography (CTG) is a cornerstone of prenatal monitoring, providing continuous fetal heart rate (FHR) recordings via Doppler ultrasound to assess fetal wellbeing. Although performed routinely (particularly in high-risk pregnancies) expert clinical interpretation is fraught with variability: sensitivity for key pathological patterns ranges from 8 % to 65 %, false positive rates remain high [1, 2], and inter-rater agreement is poor (κ = 0.12–0.39) [3-5]. These shortcomings undermine the reliability of CTG as a decision-support tool and contribute to both missed distress and unwarranted interventions.

A substantial contributor to interpretative error is signal noise, which affects a large proportion of clinical recordings [6]. Common artefacts include halving and doubling errors (causing artificial decelerations or accelerations), maternal heart rate (MHR) contamination, missing segments from transducer displacement and isolated spike anomalies [7-11]. Each noise type distorts FHR trends in unique ways, obscuring true physiological patterns or generating spurious alarms. Such artefacts have been directly implicated in delayed recognition of fetal compromise, misclassification of normal tracings, and unnecessary caesarean deliveries.

To mitigate subjectivity, rule-based systems like the Dawes-Redman™ algorithm have been widely adopted for standardised CTG analysis [12-17], and more recent efforts have explored machine learning (ML) and deep learning (DL) for FHR pattern recognition [18-23]. However, these methods assume 'clean' signals or rely on basic interpolation that addresses only missing samples and fails to correct complex artefacts such as halving/doubling errors, spike outliers or MHR overlap. As a result, their performance degrades when corrupted signals mask or mimic clinically important patterns.

In this study, we hypothesise that an end-to-end AI framework, explicitly trained to detect diverse CTG artefacts and reconstruct clean FHR traces, will enhance signal quality and improve diagnostic accuracy. We introduce *CleanCTG*, a dual-stage DL model combining multi-scale noise detection with artefact-specific reconstruction. We evaluate its effectiveness on both noise-augmented data and clinical CTG recordings, benchmarking against existing baselines and expert clinical evaluation to demonstrate its potential to improve fetal monitoring.

# Methods

## Data Preprocessing

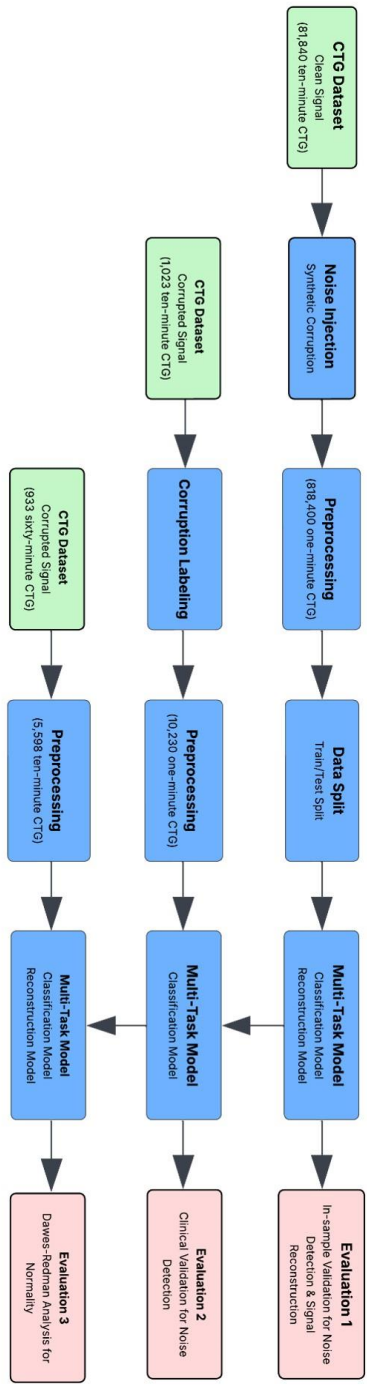

Figure 1: Data flow for the development, testing and validation of CleanCTG.

CTG recordings were obtained from the Oxford Maternity (OxMat) database, a longitudinal repository of antepartum tracings collected at the John Radcliffe Hospital, Oxford UK, between 1991 and 2024 [24]. We selected all available FHR signals sampled originally at 4Hz and applied a uniform down-sampling to 1Hz to reduce computational overhead while preserving characteristic features. Each recording, varying in length from ten minutes up to one hour, was divided into non-overlapping ten-minute segments, matching the minimum window recommended for CTG interpretation. These ten-minute windows served as the foundation for all three evaluation pipelines and ensured consistent context for noise injection and model assessment [24]. Figure 1 illustrate the preprocessing pipelines for our study.

A clean subset of 81,840 ten-minute CTG segments was selected from OxMat, containing no missing data or identifiable artefacts. Five noise types were then injected: halving errors, where every other heartbeat is dropped; doubling errors, where each beat is recorded twice; MHR contamination replacing the fetal signal; missing segments from transducer displacement or fetal movement; and isolated spike artefacts. Noise was applied randomly to each segment, corrupting up to 50% of samples, and combined scenarios were included to mimic clinical patterns (for example, MHR artefacts sandwiched between missing segments). Further details of the method are explained in Supplementary Material S1. The resulting 81,840 injected segments were further divided into 1-minute samples, yielding 818,400 segments used exclusively for the first model development and evaluation pipeline.

## External validation data

| Artefact | Synthetic CTG | | Clinical-Annotated CTG | |
|---|---|---|---|---|
| | Total Segments | Total Time (seconds) | Total Segments | Total Time, (seconds) |
| Halving | 3.39% (27,743) | 0.70% (344,688) | 0.09% (9) | 0.01% (47) |
| Doubling | 3.40% (27,825) | 0.70% (343,444) | 0.23% (23) | 0.06% (349) |
| MHR | 3.28% (26,843) | 1.02% (499,819) | 6.61% (674) | 1.24% (7594) |
| Missing | 44.10% (360,914) | 6.35% (3,118,932) | 25.53% (2,601) | 5.40% (32,993) |
| Spike | 76.00% (621,984) | 2.38% (1,167,031) | 3.97% (405) | 0.41% (2,486) |

Table 1: Proportion of synthetic noise types introduced and noises types in clinical-annotated CTG.

For external validation of the artefact detection algorithm, 1,019 ten-minute CTG recordings were independently annotated by clinicians for the presence and type of artefacts. These clinical-annotated CTG recordings were split into non-overlapping one-minute segments, yielding 10,190 samples that captured authentic noise patterns encountered in practice. No additional preprocessing (e.g. interpolation or filtering) was applied to ensure preservation of the true artefact characteristics for unbiased external validation of the detection algorithm. Table 1 summarises the proportion of each artefact in both synthetic and expert-annotated data.

For assessment of signal reconstruction, recordings of at least 60-minutes were required to perform comprehensive Dawes-Redman™ analysis, the clinical standard for CTG evaluation. Each recording was categorised into the Adverse Pregnancy Outcome or Normal Pregnancy Outcome cohort, following criteria detailed in Supplementary Material S2. Recordings with more than 50% missing data in any one-minute segment were excluded to ensure reliable reconstruction. No additional preprocessing was applied, in order to preserve the native clinical characteristics of the CTG signals. The final dataset comprised 933 recordings of 60-minutes each, corresponding to 5,598 ten-minute segments for reconstruction evaluation. The Dawes-Redman™ algorithm and the development of Adverse and Normal Pregnancy Outcome cohorts have also been detailed elsewhere [12, 14].

## Model Architecture

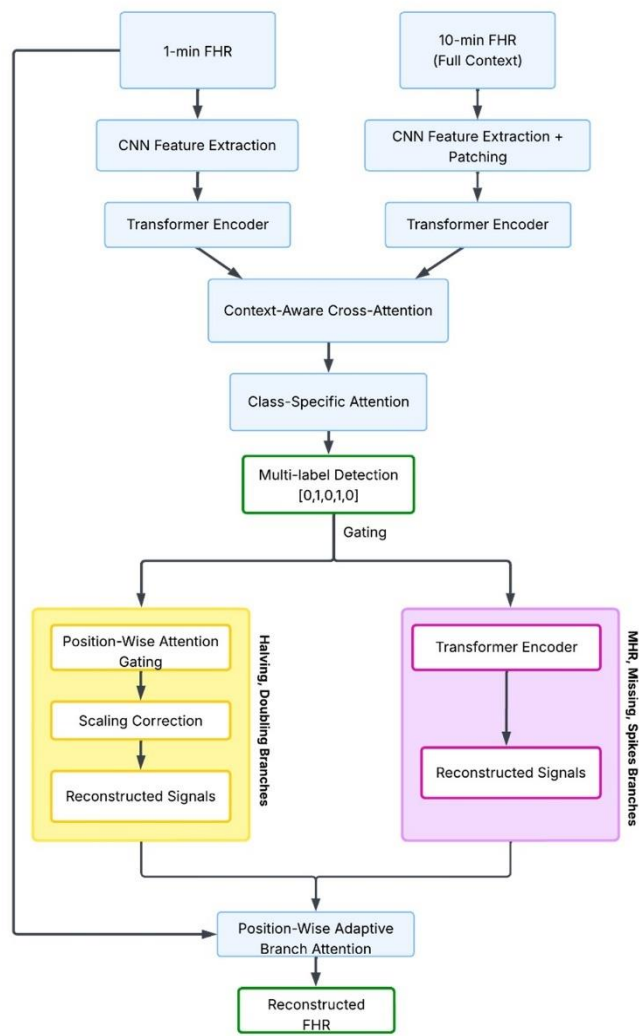

Figure 1: CleanCTG Model Architecture.

To accurately reconstruct or 'correct' the FHR in CTGs, we propose that the system must first possess the ability to correctly detect and classify artefacts present in each segment. We propose a novel approach in this field, instead of uniformly attempting to denoise all artefact types within a single unified branch. Our approach allows us to develop specialised branches intended to correct each artefact type individually, rather than using a generalised model, thereby reconstructing a more accurate representation of the original signals. To achieve this, we developed a two-stage multi-branch gating algorithm with position-wise attention. This model incorporates convolutional layers, transformer encoders, cross-attention, and class-wise attention for artefact detection, followed by specialised reconstruction and position-wise branch attention of FHR signals. Figure 2 illustrates the overall architecture of our proposed model, CleanCTG. The CleanCTG model pipeline is as follows:

1. **Feature Extraction:** CNN has been commonly used in the field of time-series biomedical processing tasks to extract features before further downstream process [22, 25-29]. In our approach, the 1-minute FHRs are passed through multiple CNN layers with varying kernel sizes to preserve sharp features. Additionally, we aim to learn features that are context-aware by considering the full 10-minute segment from which each segment was extracted. This context helps determine whether signal patterns are abnormal relative to other time points. The full segment is passed through multiple additional layers with varying kernel sizes.

2. **Self-attention Encoding:** The dual sets of CNN features are independently processed by separate transformer encoders. Each encoder consists of multiple layers comprising Layer Normalization, Multi-Head Self-Attention, and a Feedforward MLP block connected via residual connections. These enable the model to efficiently learn long-range dependencies and relationship within the feature sequences [30].

3. **Cross-Attention Transformer**: Cross-attention was applied between the 1-minute encoded features and the 10-minute encoded features, enabling the model to learn position-specific patterns while maintaining awareness of the global signal characteristics. Unlike the cross-attention mechanism originally proposed in CrossViT [31], which restricts attention to a single CLS token as the query, our approach allows all feature tokens to attend to the full set of context tokens. This enables our model to capture full temporal relationships and dependencies between local patterns and global context, rather than relying on summary representations from a single CLS token. The attention output is represented as:

$$Attention(Q, K, V) = soft\max\left(\frac{QK^T}{\sqrt{d_k}}\right)V \text{ where}$$

$$Q = f_{1mins}W_Q,$$
$$K = f_{10mins}W_K,$$
$$V = f_{10mins}W_V$$

4. **Class-Specific Attention**: The cross-attended features undergo class-specific attention pooling to weight temporal positions that are most important for each specific noise type. This process produces a condensed feature representation that captures each class's focus on the most informative parts of the signal.
5. **Multilabel Classification:** Each class-specific feature representation is then processed through a dedicated multi-layer perceptron (MLP) to produce binary outputs, which are concatenated to form a 5-dimensional multilabel prediction vector enabling simultaneous detection of multiple noise types within a single signal segment.
6. **Gating Mechanism**: Our approach implements a 2-stage gating mechanism: we use the detected multilabel outputs from the classification stage in step 5 to determine which noise types are present, where the gate of each class, $g_c$ will only be activated if it crosses a probability threshold. Within each activated branch, we then apply attention mechanisms to identify the specific positions where each artefact occurs. The denoising branches are divided into two categories:
   a. **Mathematical correction branches**: for halving and doubling artefacts, which apply multilayer-attention mechanism to identify the specific positions where scaling artefacts occur, producing position-wise masks, and is generated where $M_c = [M_{c,t} \ldots \ldots M_{c,60}] \in [0,1]$ represent the binary mask for class, $c = class$ and $t = timepoint$. These masks guide the mathematical correction applied at the detected positions, implementing signal correction through multiplication by factors of two or 0.5 to restore proper amplitude scaling.
   b. **Transformer-based branches**: For other noise types (MHR, missing, spikes), we employ vanilla transformer-based denoisers that learn end-to-end signal reconstruction. Each denoiser is activated by global detection gates:
   $$X_{reconstructed,c} = Transformer_c(X) \odot g_c + X \odot (1 - g_c)$$
   Unlike mathematical correction branches, these transformers implicitly determine denoising positions through their internal attention mechanisms, learning to reconstruct clean signals without explicit position masks.
7. **Position-wise Attention Reconstruction:** In the final combination layer, we use a position-wise attention mechanism to determine which output branches to select at each timepoint. The original signal is included as one of the branch options, serving as an implicit skip connection similar to those in CNNs or U-Net architectures. This enables the model to preserve clean signal regions by selecting the unprocessed signal when no artefacts are detected, preventing unnecessary processing that could introduce additional noise. The learned attention weights determine which signals from specific processing branches to select at each time-point, reconstructing complete signals by intelligently combining outputs produced by all branches.

## Model Training

Model training proceeded in two distinct stages. In the first stage, the artefact-detection module was trained as a multilabel classifier using binary cross-entropy loss and an adaptive optimiser with a modest learning rate and moderate mini-batch size. Once this component attained near-optimal discrimination (as measured by AU-ROC) on held-out data, its parameters were frozen.

In the second stage, the reconstruction module was trained using the frozen noise predictions as an additional input. Training minimised a composite loss (binary cross-entropy plus mean squared error) under the same optimisation regime as before. Only CTG recordings with synthetically injected artefacts were employed, and data were partitioned according to a conventional training-to-test ratio (95% training including validation and 5% test set).

## Evaluation

The evaluation phase encompassed three sequential assessments, each targeting a different level of clinical realism. First, we measured performance in a fully controlled synthetic setting; next we validated generalisation on semi-synthetic data with more complex artefact patterns; and finally, we tested on genuine clinical CTG recordings to confirm robustness and practical utility.

## Noise Detection

The artefact detection component evaluates the model's ability to classify multiple artefact types simultaneously. Given the limited availability of ML or DL approaches in existing CTG noise detection literature, which predominantly relies on rule-based methodologies, we establish comparative benchmarks using four neural network architectures: a three-layer MLP classifier, a one-dimensional residual network (ResNet), a transformer-based classification model, and a bidirectional gated recurrent unit (GRU) architecture variant based on the MHR detection algorithm, similar to the one proposed by *Boudet et al* [32]. We also adapted a variant of compact convolutional transformer (CCT), a model popular in time series biomedical analysis [33], along with TimesNet [34], a computationally efficient state-of-the-art (SOTA) general-purpose time series classifier. The hyperparameters of the comparison models are presented in Supplementary Material S3.

Classification performance was quantified using four established metrics: area under the receiver operating characteristic curve (AU-ROC), accuracy, sensitivity and specificity. To determine the best model, AU-ROC was chosen as the main metrics as it provides a threshold-independent valuation of discriminative ability. The evaluation is conducted on two distinct datasets: (i) synthetic CTGs and (ii) the clinically-annotated CTGs.

## FHR Reconstruction

FHR reconstruction evaluation assessed our model's ability to restore clean FHR from noise-contaminated recordings. We evaluated our model against five established baseline methodologies, encompassing both conventional approaches used for cleaning FHR signals and popular DL approaches commonly used to reconstruct signals from contaminated samples in biomedical time series data, as well as two popular general-purpose SOTA time-series transformer models. The traditional methods include linear interpolation and autoregressive modelling, and DL baseline approaches consists of an MLP-based autoencoder, convolutional-transformer (conv-transformer) based autoencoder, and U-Net. The general-purpose SOTA time-series transformer includes PatchTST and TimesNet. Both transformer models, have been adapted for forecasting and imputation in the past, and we adapted these models for the purpose of signal reconstruction. The architectures and hyperparameters are available in Supplementary Material S4.

To provide comprehensive evaluation, we measure mean square error (MSE) separately on the noise corrupted segments and the clean segments for the 1-minute FHRs. The noise corrupted segment MSE measures the model's ability to reconstruct artefact-contaminated signal segments, evaluating how effectively the model removes noise while restoring true FHR patterns. Conversely, the clean portion MSE assesses the algorithm's ability to preserve originally clean signal segments without unnecessary processing that could distort true FHR patterns.

We further validated our model architecture through an ablation study, comparing five ablated versions of our model. The aim of the ablation study is to demonstrate the significance of our model's components relative to other conventional techniques. Multiple components of our model were systematically removed and replaced with different architectures for comparison.

First, we replaced our dedicated mathematical branches for halving and doubling artefacts with transformer-based branches, similar to other branches in our original model, which we defined as the "all-transformer branch" configuration. Second, we removed the dedicated branches for each noise type and implemented a single unified transformer encoder. Third, we utilised a five-stacked transformer model, where the signal is progressively denoised as each noise type flows through sequential transformer layers. This architecture replaces our multi-branch approach with a predetermined sequential processing layer consisting of multiple shallow transformer denoisers, each containing a transformer encoder with layer attention heads to address one specific noise type. We then replaced our adaptive position-wise attention mechanism with two alternative architectures: a simple three-layer MLP consisting of linear layers and a bidirectional LSTM with a final linear output layer. All ablated models utilised the same predictions from the noise detection stage to evaluate specifically the effect of the reconstruction architecture modifications.

Evaluation was restricted to the synthetic injected CTGs from the initial pipeline, as the clinical FHR recordings lack ground-truth clean signals. This enables validation of the model's capacity to reconstruct clean FHR signals in controlled settings with predefined artefact profiles.

### Dawes-Redman™ Analysis

To determine whether enhanced signal quality can improve clinical workflows, we compared standard and reconstructed CTG traces using the Dawes-Redman™ computerised CTG (DR-cCTG) system. DR-cCTG is the global gold standard for automated and objective fetal wellbeing assessment. We posited that cleaner FHR signals would satisfy diagnostic criteria more rapidly, allowing shorter monitoring sessions without sacrificing accuracy.

We ran the DR-cCTG analysis on the unprocessed, artefact-contaminated recordings to establish a baseline. We then applied our reconstruction algorithm as a preprocessing step and repeated the DR-cCTG evaluation. DR-cCTG evaluates continuous CTG recordings for up to 60 minutes: an initial assessment at ten minutes, followed by reviews every two minutes until normality criteria are met, at which point monitoring may cease.

By comparing time to normality on raw versus cleaned signals, we quantified reductions in required recording length and potentially improves accuracy by, for example, removing false decelerations from MHR contamination. Our primary metrics were specificity (the proportion of recordings correctly classified as normal) and time to diagnostic conclusion. We also measured sensitivity to verify that true positive detections (cases warranting further clinical attention) were preserved. This framework assesses whether algorithmic artefact removal maintains diagnostic accuracy while improving the efficiency of computerised CTG analysis.

## Results

Across our three evaluation pipelines, we employed a total of 818,400 one-minute synthetic CTG segments for model development and internal testing, 10,190 expert-annotated one-minute segments for external noise-detection validation, and 5598 ten-minute segments drawn from 933 clinical recordings for reconstruction assessment. We benchmarked the CleanCTG framework against six alternative classifiers (MLP, ResNet, transformer, GRU, CCT and TimesNet) for artefact detection, and seven established reconstruction methods (linear interpolation, autoregressive modelling, MLP autoencoder, convolutional-transformer autoencoder, U-Net, PatchTST and TimesNet) for signal restoration. Five ablated variants of our architecture were also evaluated to isolate the contribution of each component.

## Noise Detection

| Model | Noise Injected Synthetic CTG | | | | Clinical-annotated CTG | | | |
|---|---|---|---|---|---|---|---|---|
| | AU-ROC | Sensitivity | Specificity | Accuracy | AU-ROC | Sensitivity | Specificity | Accuracy |
| CleanCTG | 1.00 | 99.90% | 99.80% | 99.50% | 0.95 | 83.44% | 94.22% | 88.83% |
| TimesNet | 1.00 | 99.57% | 97.16% | 96.65% | 0.88 | 67.50% | 94.32% | 80.91% |
| CCT | 1.00 | 99.62% | 99.59% | 98.61% | 0.90 | 66.40% | 99.45% | 82.92% |
| Transformer Classifier | 1.00 | 99.50% | 98.40% | 97.70% | 0.92 | 74.73% | 98.65% | 86.69% |
| ResNet | 1.00 | 99.60% | 98.50% | 97.10% | 0.91 | 61.41% | 96.86% | 79.13% |
| MLP Classifier | 0.99 | 99.20% | 96.10% | 92.90% | 0.91 | 68.73% | 94.67% | 81.70% |
| Bi-Directional GRU | 0.99 | 99.90% | 95.14% | 95.53% | 0.91 | 54.26% | 96.73% | 75.49% |

Table 2: Average artefact detection performance for each model. Shown are AU-ROC, sensitivity, specificity and accuracy on synthetic CTGs and on clinically annotated CTGs.

The performance of CleanCTG and baseline models is presented in Table 2. On the synthetic CTGs, all models demonstrated comparable performance with average AU-ROC scores of 1.00 or 0.99. Similarly, all models achieved average sensitivity rates exceeding 99%, with our proposed model attaining the highest specificity of 99.80%, followed by CCT at 98.59%. Our model also achieved the highest overall accuracy on the synthetic dataset at 99.50%, followed by CCT at 98.61%.

Although all models performed well on the synthetic CTGs, CleanCTG demonstrated superior generalisation to clinical-annotated CTGs, achieving the highest average AU-ROC of 0.95 on the clinical annotated dataset (p-values < 0.001), while next best performing model, transformer classifier achieved AUC of 0.92 (Table 2) CleanCTG also demonstrated the highest sensitivity at 83.44% (significantly better than the 2$^{nd}$ best model, Transformer classifier at 74.73%) and overall accuracy of 88.83% among all models. CleanCTG demonstrated a specificity of 94.22%, within 5.23% of the top performing model (CCT). When examining performance by noise type, spike artefact emerged as the most challenging category for all models, where our model significantly outperformed competing models, including SOTA models (p-values < 0.001). Our model achieved an AU-ROC of 0.77 for spike detection, representing approximately 14.93% improvement over the MLP classifier's performance. The performance of all models for all noise type is presented in Supplementary Material S5.

## FHR Reconstruction

| Model | MSE (Artefacts) ($\times 10^{-4}$) | MSE (Clean) ($\times 10^{-6}$) |
|---|---|---|
| CleanCTG | 2.74 | 2.40 |
| TimesNet | 9.09 | 42.00 |
| PatchTST | 10.80 | 91.00 |
| Conv-Transformer | 8.29 | 8.39 |
| U-Net | 9.72 | 2.11 |
| MLP Encoder | 9.69 | 200.00 |
| Autoregression | 30.52 | - |
| Linear Interpolation | 20.67 | - |

Table 3: Mean squared error for signal reconstruction methods on artefact-contaminated ($\times 10^{-4}$) and clean ($\times 10^{-6}$) CTG segments.

Table 3 presents the reconstruction performance across both noise contaminated and clean signal portions. For reconstruction of noise-contaminated segments, our proposed model achieved the lowest (best) MSE of $2.74 \times 10^{-4}$. The second-best performing model was the Conv-Transformer Encoder at $8.29 \times 10^{-4}$ followed by TimesNet at $9.09 \times 10^{-4}$. Traditional signal processing approaches performed considerably worse, with autoregression achieving $35.20 \times 10^{-4}$ and linear interpolation at $26.70 \times 10^{-4}$. For clean signal preservation, the U-Net Encoder achieved the lowest MSE of $2.11 \times 10^{-6}$, marginally outperforming our proposed model at $2.40 \times 10^{-6}$. This was followed by the Conv-Transformer Encoder at $8.29 \times 10^{-6}$.

Our model excelled compared to other methods for signal reconstruction on the synthetic dataset, demonstrating that the two-stage "detect and correct" approach outperforms alternative methodologies. Figure 3 illustrates the reconstruction performance across different artefact types on the synthetic dataset. MHR artefacts proved the most challenging to reconstruct across all models, with worse performance compared to other noise types. However, our model still proved the best in reconstructing FHR segments with the MSE for MHR artefacts at 0.009, followed by MLP and U-Net at 0.0016 and 0.0017 respectively. Notably, our model achieved near-perfect reconstruction with MSE values approaching zero for doubling artefact at $1.10 \times 10^{-6}$. When analysing performance by noise type, traditional methods such as linear regression which is commonly used in existing studies, performed worse than all DL models, with the exception of linear interpolation for spike artefacts.

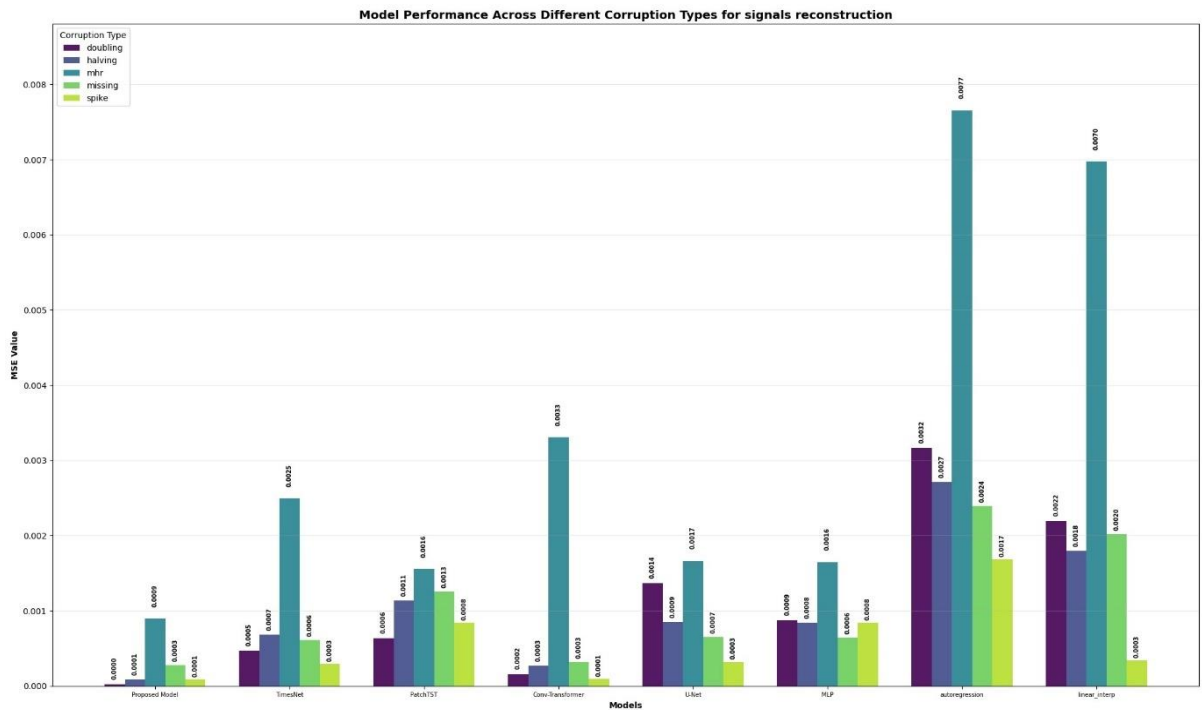

Figure 3: MSE of signal reconstruction for each model and noise type on synthetic CTG segments.

We evaluated several architectural variants by replacing our multi-branch approach which combines mathematical branches with spatial awareness and transformer encoder approaches with alternative configurations including all-transformer branches, single branches model such as a single transformer and a stacked transformer. Additionally, we ablated the position-adaptive spatial awareness mechanism, replacing it with simple MLP and LSTM alternatives.

Our proposed model, CleanCTG consistently achieved the lowest MSE for both artefact segment reconstruction ($2.74 \times 10^{-4}$) and clean segment preservation ($2.40 \times 10^{-6}$) as seen in Table 4. The poorest performing approaches utilised single-branch architectures, with the unified transformer and stacked transformer configurations performing at least 2.5× worse than our proposed approach for artefact reconstruction at $7.16 \times 10^{-4}$ and $8.69 \times 10^{-4}$ respectively. In comparison, having dedicated branches, lower MSE to $6.65 \times 10^{-4}$, as seen in All-Transformer branches model. For clean segment preservation, the second-best performing model exhibited MSE values approximately more than 2× higher than our approach at $5.00 \times 10^{-6}$, demonstrating our method's superior ability to preserve clean signal regions.

| Model | MSE (Artefacts) × 10⁻⁴ | MSE (Clean) × 10⁻⁶ |
|---|---|---|
| CleanCTG | 2.74 | 2.40 |
| All-Transformer branches | 6.65 | 8.00 |
| Unified Transformer | 7.16 | 5.00 |
| Stacked Transformer | 8.69 | 19.00 |
| Last Layer MLP | 4.70 | 32.00 |
| Last Layer LSTM | 4.89 | 6.00 |

Table 4: Mean squared error for signal reconstruction by proposed and ablated models on artefact-contaminated segments ($\times 10^{-4}$) and clean segments ($\times 10^{-6}$).

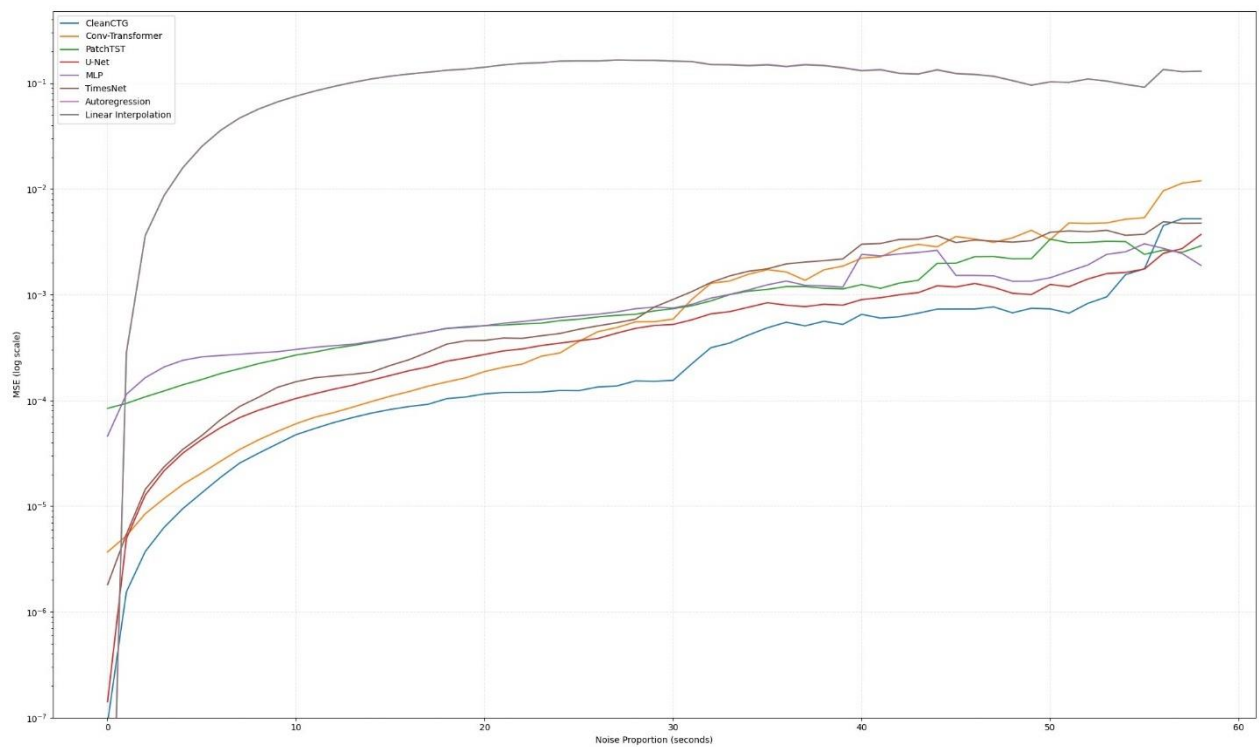

Figure 4: MSE of signal reconstruction as a function of corrupted segment length for each model

Next, we examined how varying lengths of artefact corruption affect reconstruction quality across all models. Figure 4 illustrates the impact of noise proportion for each model. Reconstruction error increased for all methods as the proportion of noise grew (from 3 to 60 timepoints). Conventional techniques (linear interpolation and autoregression) achieved comparable low MSE comparable to CleanCTG when corruption was under three seconds, but their error rose steeply thereafter. CleanCTG maintained the lowest MSE across most corruption lengths, indicating robust handling of artefact durations typical in clinical CTGs. Interestingly, when more than 90 % of a segment was corrupted, the MLP autoencoder produced lower average MSE than other DL approaches; however, visual inspection revealed excessive smoothing of the reconstructed traces, suggesting limited practical utility despite the numerical result.

## Dawes-Redman™ Analysis

| Model | Specificity (Normality) | Sensitivity | Mean | | Median | |
|---|---|---|---|---|---|---|
| | | | Time to Decision (minutes) | Improvement | Time to Decision (minutes) | Improvement |
| Raw Data | 80.70% | 40.70% | 22.8 (21.08 – 24.52) | – | 18 (10 - 32) | - |
| CleanCTG | 82.70% | 40.90% | 20.6 (18.94 – 22.18) | 9.64% | 12 (10 - 32) | 33.33% |
| TimesNet | 80.20% | 57.20% | 21.1 (19.24 – 22.49) | 7.46% | 16 (10 – 32) | 11.11% |
| PatchTST | 68.20% | 36.10% | 23.5 (12.58 – 25.55) | - 3.07% | 20 (10 - 36) | -11.11% |
| Conv-Transformer | 80.70% | 40.90% | 22.6 (20.90 – 24.40) | 0.88% | 16 (12 - 32) | 11.11% |
| U-Net | 82.40% | 35.00% | 21 (19.42 – 22.58) | 8.30% | 14 (10 - 32) | 22.22% |
| MLP | 60.60% | 71.90% | 25.2 (23.14 – 27.24) | - 11.11% | 22 (12- 38) | -22.22% |

Table 5: Dawes-Redman™ analysis results on raw and processed CTG signals, showing specificity and sensitivity, and mean and median time to decision (minutes) with percentage improvement relative to raw data.

Table 5 presents the Dawes-Redman™ analysis results comparing our proposed model against other DL-based reconstruction models and the baseline raw signal analysis. Among all reconstruction approaches, only our proposed model demonstrated slight improved normality detection performance compared to the baseline, achieving specificity scores of 82.70%.

Additionally, our proposed model-maintained sensitivity at 40.90%, closely preserving the baseline performance of 40.70%, while U-Net showed degraded sensitivity at 0.350. This indicates that our algorithm enhances the accuracy of normality detection without compromising the baseline sensitivity. Our proposed approach achieved a 9.64% and 33.33% reduction in mean and median decision time respectively, compared to the baseline Dawes-Redman™ algorithm. The reduction in decision time is the highest among all models with the next best model, U-Net achieving 8.30% and 22.22% reduction in mean and median time respectively.

# Discussion

In this study we have introduced *CleanCTG*, a two-stage deep-learning framework that first identifies multiple artefact types in cardiotocography (CTG) signals and then reconstructs clean fetal heart rate (FHR) traces. The artefact-detection stage achieved perfect discrimination on synthetic data (AU-ROC = 1.00) and excellent performance on clinician-annotated recordings (AU-ROC = 0.95, sensitivity 83.44%, specificity 94.22%, accuracy 88.83%). The reconstruction stage reduced MSE on corrupted segments to $2.74 \times 10^{-4}$ and preserved clean segments with MSE of $2.40 \times 10^{-6}$. When integrated into Dawes-Redman™ analysis on 900 continuous CTG recordings, CleanCTG increased normality detection (specificity) by 2.5% (from 80.7% to 82.7%) and reduced median decision time by 33.3%. These results confirm that explicit artefact removal and signal reconstruction can both improve detection accuracy and accelerate clinical decision-making, supporting our hypothesis that higher-quality FHR signals enable shorter monitoring sessions without loss of diagnostic fidelity.

Noise contamination in CTG traces is a well-recognised challenge that undermines the reliability of automated and manual FHR interpretation. Prior AI research has largely concentrated on applying end-to-end outcome-based prediction models to raw CTGs, with minimal attention to signal pre-processing or quality [21, 22, 35-37]. Between 50–90% of studies focus exclusively on handling missing samples through interpolation, while fewer than 10 % address key artefact classes such as maternal heart rate interference [38]. Rule-based methods and basic signal-processing algorithms have limited adaptability to diverse artefact phenotypes or durations [18, 23, 39-41]. A bidirectional GRU network incorporating both FHR and MHR inputs achieved 93.1 % sensitivity for maternal-heart-rate detection but fell to 69.9 % when MHR data were unavailable [32]. Another bespoke outlier-removal network doubled the F1-score for suspicious case prediction but did not reconstruct the underlying FHR [42]. However, no existing approach provides a unified pipeline for comprehensive artefact detection and artefact-specific signal correction.

CleanCTG addresses these gaps through three methodological advances:

1. Dual-Scale Convolution and Context-Aware Cross-Attention: We extract features at multiple scales via separate convolutional backbones, then align local and global representations using a cross-attention mechanism that allows each local token to attend to all contextual tokens. This design improves sensitivity to both brief and extended artefacts without relying on a single summary token.

2. Hierarchical Gating with Artefact-Specific Reconstruction Branches: A multilabel classification stage produces artefact presence gates. Scaling errors (halving and doubling) are corrected via mathematically defined branches that apply position-wise masks and amplitude adjustments. Other artefact types (MHR overlap, missing segments, spikes) are handled by dedicated transformer-based denoisers. A final position-wise attention layer weights outputs from each branch and the original signal, preserving uncorrupted regions while applying targeted corrections.
3. Systematic Synthetic-Noise Injection Protocol: We generated five physiologically-driven realistic artefact types (halving, doubling, MHR overlap, missing segments and isolated spikes), applying them randomly and in combination to clean CTG segments. This approach yielded 818,400 one-minute samples for initial training and ensured robustness to a broad range of clinical noise patterns in the absence of large-scale annotated data.

In the first evaluation pipeline using synthetic data, CleanCTG achieved perfect artefact detection (AU-ROC = 1.00) and delivered exceptionally low reconstruction error on corrupted segments (MSE = $2.74 \times 10^{-4}$), while preserving uncorrupted traces with minimal over-processing (MSE = $2.40 \times 10^{-6}$). In the second pipeline, which employed 10,190 expert-annotated one-minute segments, CleanCTG outperformed six alternative neural classifiers (MLP, ResNet, transformer, GRU, compact convolutional transformer and TimesNet), achieving the highest AU-ROC 0.95. Finally, under the Dawes-Redman™ system in the third pipeline, applying CleanCTG to clinical CTG recordings increased normal-pattern specificity by 2.5% without reducing sensitivity and shortened the median time to decision by 33.3 %, signalling potential for fewer unnecessary interventions and more efficient monitoring. In contrast, general-purpose SOTA time-series models (TimesNet, PatchTST) struggled. TimesNet's AU-ROC fell to 0.88 on clinical data and to 0.50 for spike detection, while PatchTST showed consistently inferior MSE across all categories. Such results highlight that SOTA architectures tuned for forecasting or imputation may underperform on domain-specific biomedical tasks when lacking artefact-aware design.

Our ablation study showed that having multiple artefact specific branches along with position-wise attention improved the ability for signal reconstruction, compared to single branches variants. This is evidenced in Table 4, where Unified Transformer and Stacked Transformer achieved the highest MSE compared to those with dedicated branches. In addition, our ablation study also confirmed that the position-wise attention mechanism is critical for preserving clean segments. Replacing it with three-layer MLP or bidirectional LSTM reduced signal fidelity, whereas CleanCTG matched U-Net's performance in conserving noise-free regions without requiring multiple skip connections.

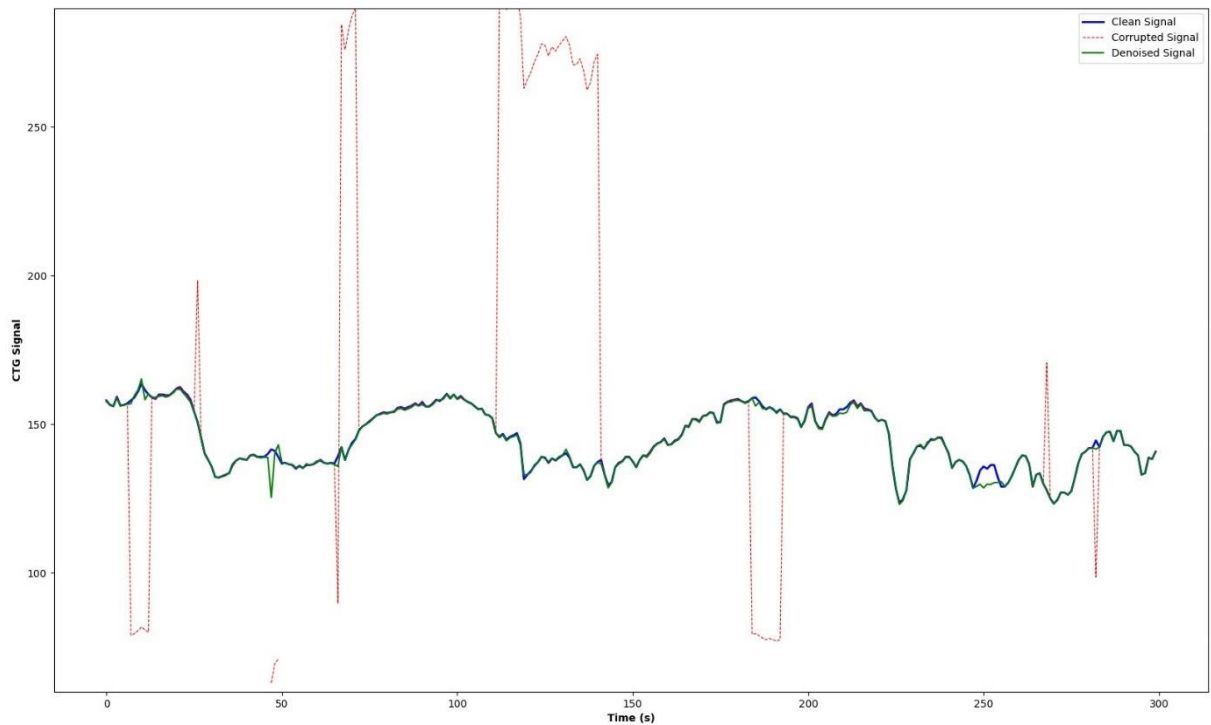

Figure 5: Reconstruction of a 5 minutes synthetic corrupted CTG with CleanCTG.

CleanCTG's artefact-specific reconstruction demonstrates clear advantages in real-world scenarios. In cases of halving and doubling errors, the model applies precise mathematical corrections only at the detected timepoints, restoring true FHR without altering adjacent clean segments (see Figure 5 and Supplementary Material S6 & S7 Figure). By contrast, standard U-Net reconstruction often fails to correct scaling artefacts and can introduce spurious accelerations, while general-purpose transformers such as TimesNet not only overlook doubling errors but also generate negative spikes, highlighting the need for dedicated correction logic.

Equally important is CleanCTG's ability to preserve critical clinical events. As shown in Supplementary S8 Figure, the model correctly differentiates genuine fetal bradycardia from maternal-heart-rate contamination, retaining the true bradycardia pattern for accurate Dawes-Redman™ interpretation. In rare instances, brief accelerations may be misclassified as spikes and smoothed, leading to minor delays in reaching normality criteria; however, these delays do not affect the final clinical decision regarding fetal wellbeing. These use cases illustrate CleanCTG's potential to enhance diagnostic reliability and efficiency in clinical CTG analysis.

Despite high AU-ROC, sensitivity for certain artefacts declined when using fixed probability thresholds on clinical data, indicating a domain shift from synthetic to real signals. Addressing this will require larger volumes of expert-annotated clinical CTGs to optimise gate thresholds. Furthermore, the clinical annotation set remains small relative to synthetic data. Manual annotation is time-intensive and demands specialised expertise. We also did not evaluate inference speed or resource requirements in live clinical environments; these practical considerations will be critical for real-time deployment.

To enhance robustness and clinical utility, future work will:

- Expand clinical annotation efforts to refine threshold calibration and improve sensitivity, especially for spike artefacts.
- Conduct prospective studies in labour wards to validate performance under real-time monitoring conditions.
- Explore model compression, pruning or knowledge distillation techniques to reduce computational overhead, enabling integration into resource-constrained monitoring systems.
- Investigate transfer learning across related biomedical time-series tasks to assess adaptability of the CleanCTG architecture.

# Conclusion

CleanCTG represents the first end-to-end deep-learning solution that detects a comprehensive set of CTG artefacts and reconstructs clean FHR signals. Through novel dual-scale attention mechanisms, hierarchical gating and artefact-specific correction branches, the model demonstrates superior performance in synthetic, semi-synthetic and clinical evaluations and yields meaningful improvements in automated CTG analysis. By improving signal quality and reducing decision-making time, CleanCTG has the potential to enhance the reliability and efficiency of fetal monitoring in clinical practice.

# Acknowledgements

The authors wish to thank Ravi Shankar for his assistance with data management.

# Declaration of competing interest

# Data availability statement

The dataset used in this study is not available publicly due to privacy and ethical considerations. However, additional information of the dataset can be found here [24].

# Funding

This research was supported by the UKRI Medical Research Council (MR/X029689/1).

## Author contributions

**Sheng Wong**: Writing – original draft, Methodology, Formal analysis, Validation & Visualization; **Beth Albert**: Project administration, Validation, Writing – review and editing; **Gabriel Davis Jones**: Conceptualization, Methodology, Validation, Writing – review and editing & Supervision.

**S1. Description of the Noise Generation Process**

The noise generation process involved systematically injecting corrupted segments into corruption-free 10-minute CTG recordings. Each corruption type was implemented to accurately replicate real-world artifacts observed in clinical practice. For halving and doubling artifacts, we selected random-length segments and modified the signal values by either halving or doubling them to reflect authentic examples of these monitoring errors. MHR artifacts were simulated by replacing random segments with synthesized MHR patterns ranging between 70 - 110 bpm. Missing signal segments were created by selecting random-length chunks and replacing them with null values to simulate gaps caused by fetal repositioning or transducer displacement. Spike artifacts were generated by introducing abrupt changes of 5 - 40 bpm in both positive and negative directions, randomly distributed throughout the signal. We incorporated compound corruption patterns that mirror real-world scenarios. Missing value segments were injected before and after halving, doubling, and MHR artifacts to replicate the multiple simultaneous corruptions frequently observed in clinical settings.

The injection probabilities were calibrated based on clinical expertise and real-world occurrence rates. Doubling and halving artifacts were assigned a 5% injection probability each, reflecting their relative rarity in clinical practice while acknowledging their serious diagnostic implications when misinterpreted. Other artifact types received a 10% injection probability to better represent their higher chance of occurring in CTGs recording. We also included several constraints to our synthetic noise generation. The order of noise injection type was randomized, and total corruption was limited to 50% of each segment to preserve sufficient clean signal for meaningful analysis. Additionally, continuous noise segments for each artifact type were limited to 5% of the total segment length to prevent any single corruption type from dominating the entire recordings.

**S2: Classification Criteria for Adverse Pregnancy Outcome cohort and Normal Pregnancy Outcome cohort**

Adverse Pregnancy Outcome cohort and Normal Pregnancy Outcome cohort criteria were used in our study during the third evaluation phase to categorize CTG recordings based on accompanying medical information. These recordings enable evaluation of Dawes-Redman analysis effectiveness with and without our proposed reconstruction algorithm. The classification criteria are outlined below:

Adverse Pregnancy Outcome cohort:

- Birthweight 3rd percentile or below with 1-minute Apgar score less than 4 and 5-minute Apgar score less than 7
- Acidemia without labour: pH less than 7.13 and base excess greater than 10
- Acidemia with labour: pH less than 7.05 and base excess greater than 14
- 1-minute Apgar less than 4 and 5-minute Apgar less than 7

- Stillbirth before labour
- Stillbirth during labour
- Other stillbirth
- Death within 24 hours of birth
- Neonatal death
- Early neonatal death
- Asphyxia
- Hypoxic-ischemic encephalopathy
- Confirmed neonatal sepsis
- Perinatal infections
- Respiratory conditions
- 7 or more days in neonatal intensive care unit
- 7 or more days in special care baby unit

Normal Pregnancy Outcome cohort:

- Maternal age 18-39 years
- Maternal BMI 30 or less
- Delivery between 37-41 weeks gestation
- Live birth
- Birthweight between 10th-90th percentiles
- 1-minute Apgar score 4 or higher
- 5-minute Apgar score 7 or higher
- No resuscitation required
- No neonatal intensive care unit days
- No special care baby unit days
- No intensive therapy unit admission
- No perinatal infections
- No respiratory conditions

**S3. Hyperparameters used for the comparison models in Noise Detection.**

The 3-layers MLP model serve as the baseline for all noise detection. It consists of 3 hidden layers of 512, 256, 128 neurons respectively, where each layer Is followed by batch normalisation, ReLU activation and dropout regularization of 0.1. The ResNet model was amended to suit 1D time-series CTGs. It consists of 4 convolutional layers with filters of 64, 128, 256, and 512 respectively, with a stride of 2 for temporal down-sampling. The model incorporates residual blocks with kernel size 3 and batch normalization throughout. An initial convolutional layer uses a kernel size of 15 to capture temporal patterns, followed by the residual layers. Adaptive global average pooling and dropout regularization of 0.1 were used before the final classification layer. The transformer classifier was implemented with 3 transformer encoder layers, 4 attention heads and a dimension of 256 with positional embeddings. Each layer includes multi-head self-attention followed by feed-forward networks of 256 hidden units. The model ends with global average pooling across the sequence dimension, followed by the classification head. For bidirectional GRU, a 3-layer architecture with hidden size of 64 units for each layer that processes sequences in both directions simultaneously was incorporated, where the total representation size is 128 dimensions. Dropout regularization of 0.1 was also implemented throughout the network. Average pooling was used to aggregate temporal features, and a multi-layer classification head was used for classification with dimensions of 64 and 32 before producing the final predictions.

For comparison with more commonly used SOTA architectures, CCT and TimesNet were chosen. We adapted a variant of CCT with a 3-layer CNN frontend using progressively increasing channels of 16, 32 and 64 with kernel size of 3 to preserve the sequence length. The CNN features are then tokenized into 1D patches and projected to the transformer dimension. The hyperparameters used here are similar to the transformer model and are identical, up until the classification head. As for TimesNet, all hyperparameters were defaulted to the original paper with a few exceptions to improve computational efficiency. The model utilised Fast Fourier Transform (FFT) to identify the dominant periods in the time series, then reshapes the 1D temporal variations into 2D tensors based on these discovered periods. Each times block processes these 2D representations through Inception blocks with kernel sizes of 2 to capture temporal patterns. The architecture consists of 3 times block with a model dimension of 32, feed-forward dimension of 64, and uses adaptive aggregation to combine period-specific representations. The model ends with global a flattening layer and a linear classification head.

### S4. Hyperparameters used for the comparison models in Signal Reconstruction.

The MLP Autoencoder consists of 5 fully connected layers with an encoder-decoder structure with dimensions of 128, 256, 512, 256, 128 by first learning an expanded latent representation and then reconstructing the denoised signal. Each layer incorporates GELU activation functions and dropout regularization of 0.1. The Conv-Transformer implements an encoder-decoder style architecture for signal reconstruction tasks. The model uses convolutional patch embedding with patch size 5 and embedding dimensions of 256. The core architecture consists of 6 transformer blocks with 8 attention heads and 64 dimensions per head, with pre-layer normalization and standard feed-forward networks with GELU activation. The reconstruction head converts patch embeddings back to the original signal dimensions through linear projection. A residual connection is used between the input and reconstructed output. For U-Net, it learns to predict the corruption components, then subtracts the noise components from the noisy signal to produce the clean signals. The model adapted to 1D time series uses a 4-layer deep U-Net structure with base filters of 5, progressively doubling channel dimensions through the encoder path of 32, 64, 128, 256 before contracting symmetrically in the decoder. Each layer consists of double convolution blocks with batch normalization,

ReLU activation, and dropout regularization at 0.1. The encoder utilizes max pooling for down-sampling, while the decoder employs transpose convolution for up-sampling with skip connections.

For PatchTST, which has been used for forecasting, classification and imputations, we adapted it for signal reconstruction for CTGs. The model utilizes patch embedding with a patch length of 16 to process input sequences. The architecture consists of 3 transformer encoder layers with 8 attention heads, a model dimension of 64, and feed-forward dimensions of 128. The system employs self-attention mechanisms with GELU activation functions and maintains a dropout rate of 0.1 throughout the network for regularization. In the final layer, a fusion component consisting of convolutional layers was added to preserve signal integrity for signal reconstruction purposes, combining the original input with the transformer output through a multi-layer convolutional network with dimensions of 32, 16, and 1.

For TimesNet, which has a similar purpose as PatchTST, we adapted it for signal reconstruction for CTGs. To improve computational efficiency, we maintained the hyperparameters of the original work, with a few exceptions. The forecasting approach predicts sequences of the same length as the input for reconstruction purposes, where the predicted segment represents the cleaned signal output. The architecture consists of 2 times blocks with a model dimension of 32 and feed-forward dimension of 128, utilizing Inception blocks with 3 kernel configurations for multi-scale pattern extraction. Similarly, a fusion layer incorporating three convolutional layers with dimensions 32, 16, and 1 combines the original input with the TimesNet prediction to produce the final reconstructed signal.

## S5. Performance of all models on each noise type

| Noise Type | Model | Noise Injected Synthetic CTG | | | | Clinical annotated CTG | | | |
|---|---|---|---|---|---|---|---|---|---|
| | | AU-ROC | Sensitivity | Specificity | Accuracy | AU-ROC | Sensitivity | Specificity | Accuracy |
| Halving | CleanCTG | 1.00 | 99.90% | 99.50% | 99.70% | 1.00 | 100.00% | 99.17% | 99.58% |
| | TimesNet | 0.99 | 98.65% | 99.37% | 99.01% | 1.00 | 66.67% | 99.88% | 83.27% |
| | CCT | 1.00 | 99.50% | 99.38% | 99.44% | 0.99 | 66.67% | 99.99% | 83.33% |
| | Transformer Classifier | 1.00 | 99.50% | 98.80% | 99.10% | 1.00 | 100.00% | 99.90% | 99.95% |
| | ResNet | 1.00 | 95.70% | 99.20% | 97.40% | 0.97 | 22.22% | 99.61% | 60.91% |
| | MLP Classifier | 0.99 | 90.60% | 97.60% | 94.10% | 0.97 | 66.67% | 97.29% | 81.98% |
| | Bi-Directional GRU | 0.98 | 90.60% | 99.40% | 94.40% | 0.99 | 22.22% | 99.76% | 60.99% |
| Doubling | CleanCTG | 1.00 | 100.00% | 100.00% | 100.00% | 1.00 | 91.30% | 98.75% | 95.03% |
| | TimesNet | 1.00 | 1.00% | 97.65% | 98.82% | 0.99 | 96.65% | 92.38% | 94.01% |
| | CCT | 1.00 | 100.00% | 99.98% | 99.99% | 1.00 | 95.65% | 100.00% | 97.83% |
| | Transformer Classifier | 1.00 | 100.00% | 100.00% | 100.00% | 1.00 | 95.65% | 99.92% | 97.79% |
| | ResNet | 1.00 | 99.90% | 100.00% | 99.90% | 1.00 | 95.65% | 99.70% | 97.67% |
| | MLP Classifier | 1.00 | 98.40% | 100.00% | 99.20% | 1.00 | 91.30% | 99.83% | 95.57% |
| | Bi-Directional GRU | 0.99 | 98.35% | 100.00% | 99.20% | 0.97 | 43.48% | 100.00% | 71.74% |
| Maternal Heart Rate (MHR) | CleanCTG | 1.00 | 96.00% | 99.70% | 97.80% | 0.97 | 96.44% | 82.57% | 89.51% |
| | TimesNet | 0.99 | 86.47% | 98.55% | 92.51% | 0.95 | 69.14% | 99.11% | 84.12% |
| | CCT | 0.99 | 88.95% | 99.27% | 94.11% | 0.97 | 68.84% | 98.44% | 83.64% |
| | Transformer Classifier | 0.99 | 89.50% | 97.10% | 93.30% | 0.97 | 71.51% | 98.10% | 84.81% |
| | ResNet | 0.99 | 91.40% | 98.50% | 94.90% | 0.97 | 68.84% | 98.48% | 83.66% |
| | MLP Classifier | 0.98 | 84.90% | 97.20% | 91.10% | 0.96 | 57.86% | 97.92% | 77.89% |
| | Bi-Directional GRU | 0.99 | 90.83% | 97.40% | 94.10% | 0.97 | 84.27% | 97.16% | 90.72% |
| Missing | CleanCTG | 1.00 | 100.00% | 100.00% | 100.00% | 0.98 | 95.89% | 99.83% | 97.86% |
| | TimesNet | 0.99 | 100.00% | 91.06% | 95.53% | 0.98 | 96.42% | 89.66% | 93.04% |
| | CCT | 1.00 | 100.00% | 100.00% | 100.00% | 0.97 | 95.89% | 99.82% | 97.85% |
| | Transformer Classifier | 1.00 | 100.00% | 100.00% | 100.00% | 0.97 | 95.89% | 99.83% | 97.86% |
| | ResNet | 1.00 | 100.00% | 100.00% | 100.00% | 0.98 | 95.89% | 99.82% | 97.85% |
| | MLP Classifier | 1.00 | 100.00% | 99.90% | 99.90% | 0.98 | 95.69% | 99.80% | 97.75% |
| | Bi-Directional GRU | 1.00 | 100.00% | 100.00% | 100.00% | 0.98 | 95.89% | 99.83% | 97.86% |
| Spike | CleanCTG | 1.00 | 100.00% | 99.60% | 99.80% | 0.77 | 33.58% | 90.76% | 62.17% |
| | TimesNet | 1.00 | 99.90% | 94.86% | 97.38% | 0.50 | 9.63% | 90.60% | 50.11% |
| | CCT | 1.00 | 99.84% | 99.21% | 99.53% | 0.60 | 4.94% | 99.02% | 60.34% |
| | Transformer Classifier | 1.00 | 99.60% | 92.30% | 95.90% | 0.64 | 10.62% | 95.47% | 53.04% |
| | ResNet | 1.00 | 99.90% | 86.50% | 93.20% | 0.62 | 24.44% | 86.70% | 55.57% |
| | MLP Classifier | 0.98 | 99.80% | 60.60% | 80.20% | 0.67 | 32.10% | 78.51% | 55.30% |
| | Bi-Directional GRU | 0.99 | 99.80% | 79.00% | 95.50% | 0.66 | 25.43% | 86.89% | 56.16% |
| Average | CleanCTG | 1.00 | 99.90% | 99.80% | 99.50% | 0.95 | 83.44% | 94.22% | 88.83% |
| | TimesNet | 1.00 | 99.57% | 97.16% | 96.65% | 0.88 | 67.50% | 94.32% | 80.91% |
| | CCT | 1.00 | 99.62% | 99.59% | 98.61% | 0.90 | 66.40% | 99.45% | 82.92% |
| | Transformer Classifier | 1.00 | 99.50% | 98.40% | 97.70% | 0.92 | 74.73% | 98.65% | 86.69% |
| | ResNet | 1.00 | 99.60% | 98.50% | 97.10% | 0.91 | 61.41% | 96.86% | 79.13% |
| | MLP Classifier | 0.99 | 99.20% | 96.10% | 92.90% | 0.91 | 68.73% | 94.67% | 81.70% |

**S6. Reconstructions of a 5-minutes synthetically corrupted CTG across all evaluated models. The corrupted CTG contains MHR artefacts, doubling artefacts, missing segments, and spike artefacts. The average MSE of the reconstructed of these segments are $5.10 \times 10^{-5}$ for our model, $7.60 \times 10^{-5}$ for TimesNet, $2.10 \times 10^{-4}$ for PatchTST, $5.60 \times 10^{-5}$ for Conv-Transformer, $2.73 \times 10^{-4}$ for U-net and $8.50 \times 10^{-4}$ for MLP Encoder.**

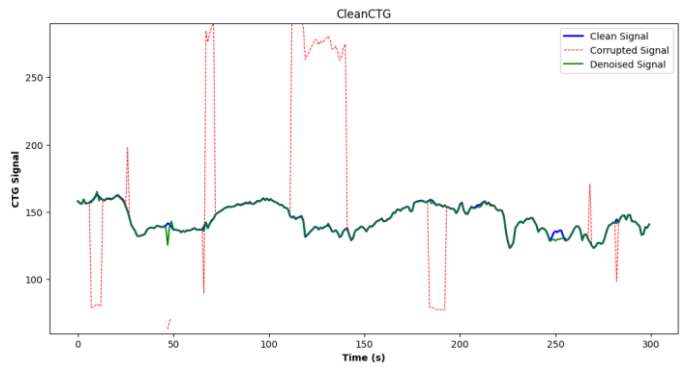
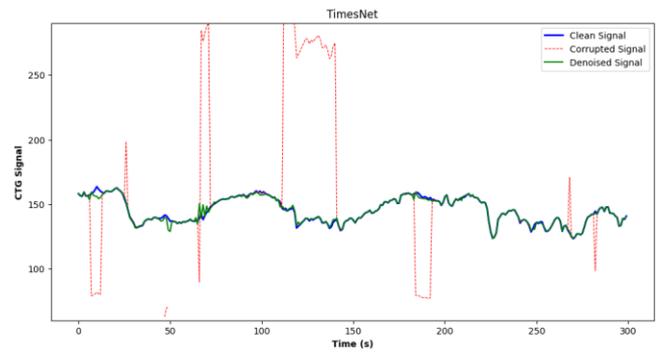
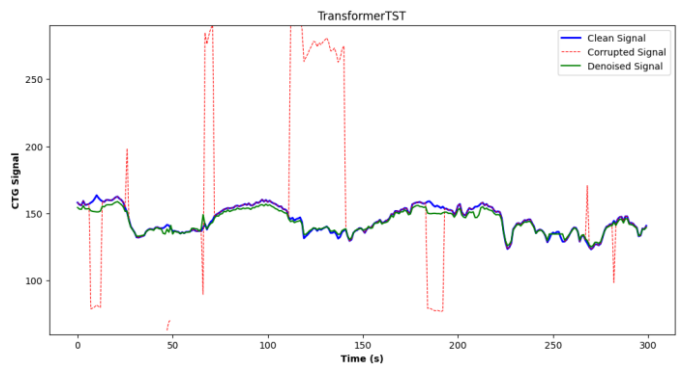
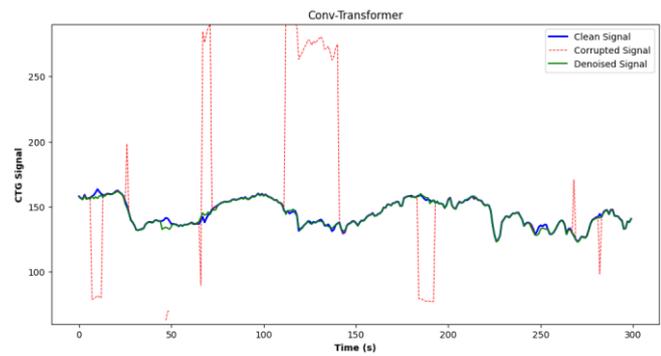
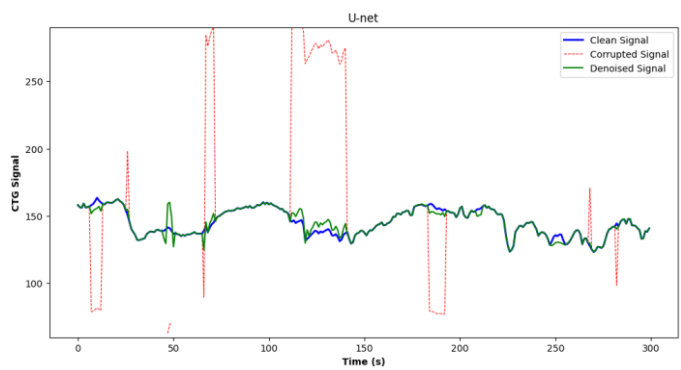
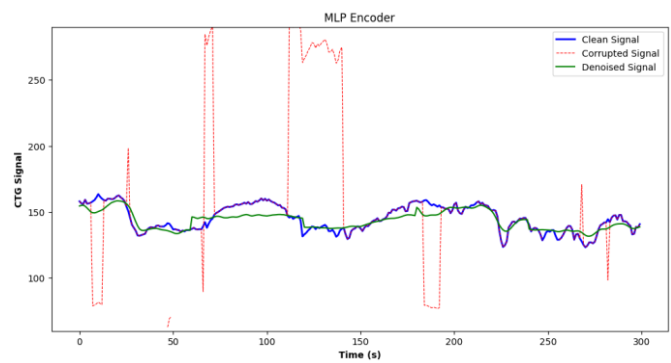

**S7. Reconstructions of a 5-minutes synthetically corrupted CTG across all evaluated models. The corrupted CTG contains MHR artefacts, doubling artefacts, missing segments, and spike artefacts. The average MSE of the reconstructed of these segments are $1.10 \times 10^{-4}$ for our model, $2.19 \times 10^{-5}$ for TimesNet, $8.50 \times 10^{-5}$ for PatchTST, $9.30 \times 10^{-5}$ for Conv-Transformer, $1.86 \times 10^{-4}$ for U-net and $2.66 \times 10^{-5}$ for MLP Encoder.**

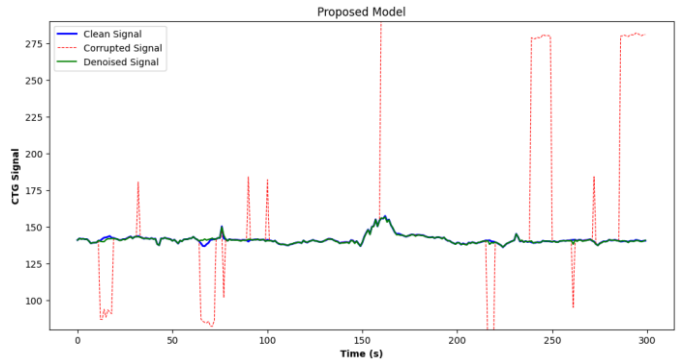
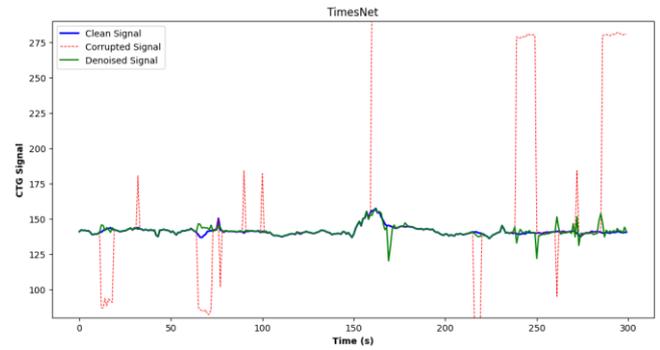
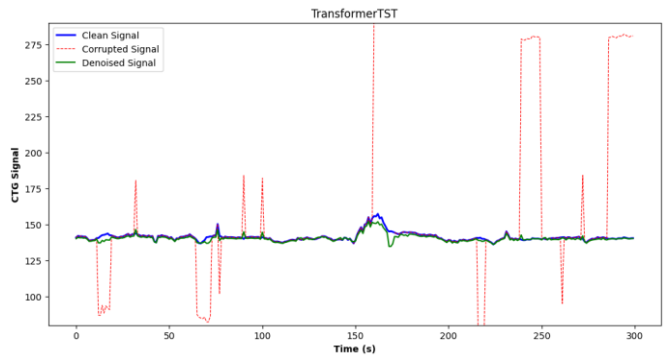
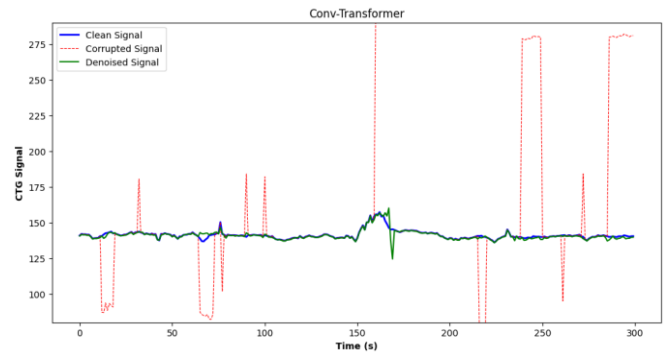
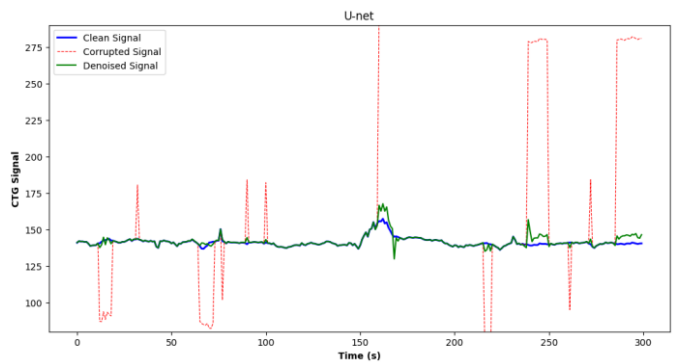
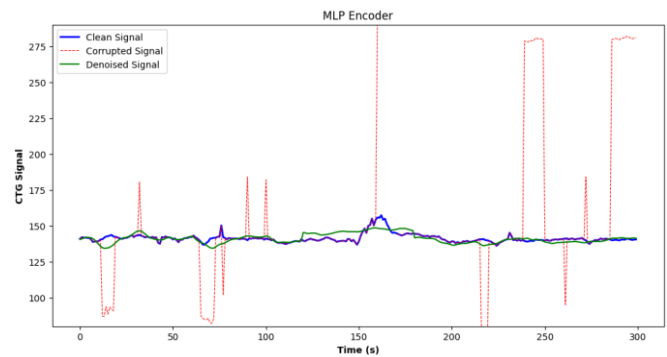

**S8. Reconstructed FHR of a CTG recordings with suspected bradycardia based on Dawes-Redman™ Analysis, plotted with a standardised plotting tool for CTG analysis. The reconstructed FHR was shifted by 10 bpm to improve visualisation.**

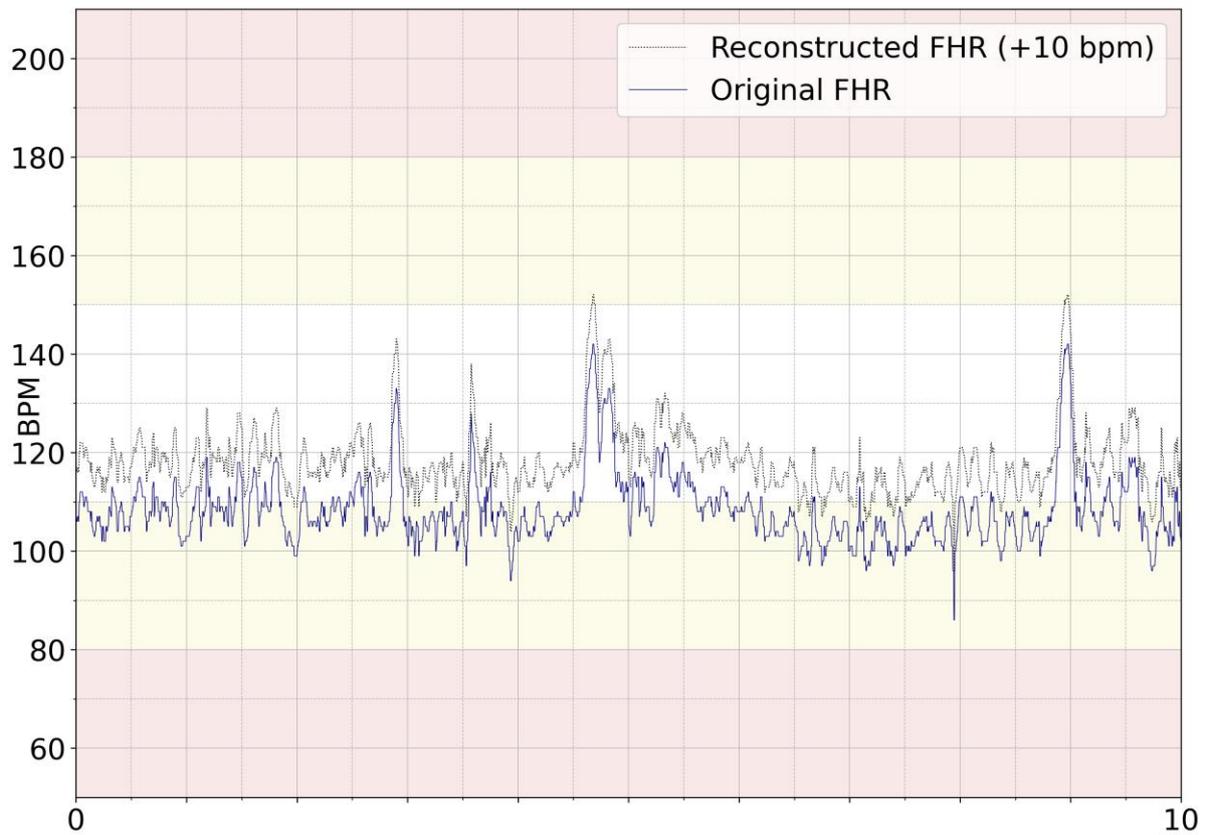